\documentclass{article}

\usepackage[letterpaper, textwidth=6.5in, textheight=9in, left=1in, right=1in, top=1in, bottom=1in]{geometry}

\usepackage{graphicx}  
\usepackage{amsmath}   
\usepackage{amssymb}   
\usepackage{calc}      
\usepackage{cite}

\title{Approximating strange attractors and Lyapunov exponents\\of delay differential equations using Galerkin projections}

\author{\ \\[-8pt]
        Anwar~Sadath\\
        {\small{Department of Mechanical and Aerospace Engineering}}\\
        {\small{Indian Institute of Technology Hyderabad, Kandi, Sangareddy, Telangana 502285, India}}\\[6pt]
        \and
        Thomas~K.~Uchida\\
        {\small{Department of Mechanical Engineering, University of Ottawa}}\\
        {\small{Colonel By Hall, 161 Louis Pasteur, Ottawa, Ontario, K1N 6N5, Canada}}\\[6pt]
        \and
        C.P.~Vyasarayani\thanks{Address all correspondence to this author.}\\
        {\small{Department of Mechanical and Aerospace Engineering}}\\
        {\small{Indian Institute of Technology Hyderabad, Kandi, Sangareddy, Telangana 502285, India}}\\
        {\small{Tel.: 040-2301-7070,~~Fax: 040-2301-6032,~~Email: vcprakash@iith.ac.in}}\\
        \ \\[6pt]}

\date{October 1, 2018}
\begin{document}
\maketitle

\begin{abstract}
Delay differential equations (DDEs) are infinite-dimensional systems, so even a scalar, unforced nonlinear DDE can exhibit chaos.
Lyapunov exponents are indicators of chaos and can be computed by comparing the evolution of infinitesimally close trajectories.
We convert DDEs into partial differential equations with nonlinear boundary conditions, then into ordinary differential equations (ODEs) using the Galerkin projection.
The solution of the resulting ODEs approximates that of the original DDE system; for smooth solutions, the error decreases exponentially as the number of terms used in the Galerkin approximation increases.
Examples demonstrate that the strange attractors and Lyapunov exponents of chaotic DDE solutions can be reliably approximated by a smaller number of ODEs using the proposed approach compared to the standard method-of-lines approach, leading to faster convergence and improved computational efficiency.
\end{abstract}

\section{Introduction}\label{sec:intro}
Nonlinear differential equations \cite{nayfeh1995applied} are widely used to model a variety of physical systems.
In the presence of transport or communication delays, these equations take the form of nonlinear delay differential equations (DDEs) \cite{balachandran2009ddes}.
Systems in which nonlinear DDEs are encountered include mathematical models of manufacturing processes \cite{balachandran2001milling,kalmarnagy2001tool,kalmarnagy2009stability}, control systems \cite{sun2004feedback,guan2007feedback,park2005feedback,erneux2007crane,wen2017direct,dong2017quadrature,zhu2016observer}, traffic flow \cite{safonov2000traffic,orosz2006car}, biological systems \cite{verdugo2008hopf,verdugo2008manifold}, population dynamics \cite{kuang1993population,kubiaczyk2002population}, shimmy dynamics \cite{takacs2009shimmy}, fluid elastic instability in heat-exchanger tubes \cite{paidoussis1992vibrations,debedout1999crossflow}, and neural networks \cite{belair1996neural,campbell1999neural}.
Like partial differential equations (PDEs), DDEs are infinite-dimensional systems.
In fact, any DDE can be equivalently posed as an abstract Cauchy problem (a PDE) with a boundary condition that is linear if the DDE is linear and nonlinear if the DDE is nonlinear \cite{maset2003cauchy,bellen2000cauchy}.

Many nonlinear DDEs exhibit chaos.
If an attractor exists, we are often interested in computing its Lyapunov exponents.
The Lyapunov exponents quantify the rate of exponential divergence (if positive) or convergence (if negative) of nearby trajectories on an attractor; a Lyapunov exponent of zero indicates neutral stability along the flow.
An $n$-dimensional system has $n$ Lyapunov exponents; the solution is chaotic if at least one Lyapunov exponent is positive.

Lyapunov exponents are calculated by monitoring the long-term evolution of an infinitesimal perturbation from a reference trajectory.
In the case of ordinary differential equations (ODEs), we typically compute the Lyapunov exponents by integrating the variational equations \cite{nayfeh1995applied,shimada1979ergodic,wolf1985lyapunov}; however, DDEs are infinite dimensional and, therefore, have an infinite number of Lyapunov exponents.
All existing work on calculating the Lyapunov exponents of DDEs has involved approximating the DDEs using finite-dimensional systems of ODEs \cite{farmer1982chaotic,mensour1998synchronization,sprott2006henon,sprott2007simple}.
A popular strategy for converting DDEs into ODEs is the method of lines (MoL) \cite{farmer1982chaotic,mensour1998synchronization}, also referred to as iterative mapping \cite{sprott2006henon,sprott2007simple} and continuous time approximation \cite{sun2009continuous}.
In the MoL approach, the domain of interest is discretized into nodes and the spatial derivatives are approximated at these nodes using finite differences; the temporal derivatives are unaltered.
The result is a large system of ODEs that approximates the original DDE.
It is well known that such finite difference--based approximations require a large number of ODEs---and, therefore, a large amount of computation time---to precisely capture the solution of a given DDE \cite{boyd2001chebyshev}.
Computational expense is of particular importance when computing Lyapunov exponents, which can require simulations of long duration.

In this work, we use a transformation \cite{maset2003cauchy,bellen2000cauchy} to convert DDEs into PDEs with nonlinear boundary conditions.
These PDEs are then converted into ODEs using the Galerkin approximation \cite{vyasarayani2012galerkin,wahi2005vibrations}.
We use Legendre polynomials as global basis functions, which allows us to obtain closed-form expressions for the integrals in the Galerkin approximation \cite{ahsan2015galerkin}.
The nonlinear boundary conditions are incorporated into the Galerkin approximation using the tau method \cite{vyasarayani2014spectral}.
The solution of the resulting ODEs approximates the solution of the original DDE system.
Provided the solution of the DDE is smooth, the Galerkin approximation guarantees that the error between the approximate and actual solutions decreases exponentially as the number of terms in the Galerkin approximation increases \cite{breda2016pseudospectral}.
As we will demonstrate with numerical examples, an ODE system obtained using the Galerkin approximation has substantially lower error than a system of the same dimension obtained using the MoL approach.
The focus of the present work is to explore the efficacy of computing Lyapunov exponents and approximating the strange attractors of nonlinear DDEs \cite{saltzman1962convection} using systems of ODEs obtained via Galerkin approximation.

This paper is structured as follows.
In Section \ref{sec.math}, we present the mathematical details for converting DDEs into ODEs using the method of lines and the Galerkin projection.
We then discuss how to compute the Lyapunov exponents of ODEs.
In Section \ref{sec.results}, we present numerical results for several nonlinear DDEs, including approximations of strange attractors and Lyapunov exponents.
Finally, we provide conclusions from this study in Section \ref{sec.conclusions}.

\section{Mathematical Modeling}\label{sec.math}
\begin{figure}[t]
    \begin{center}
        \includegraphics[width=0.55\textwidth]{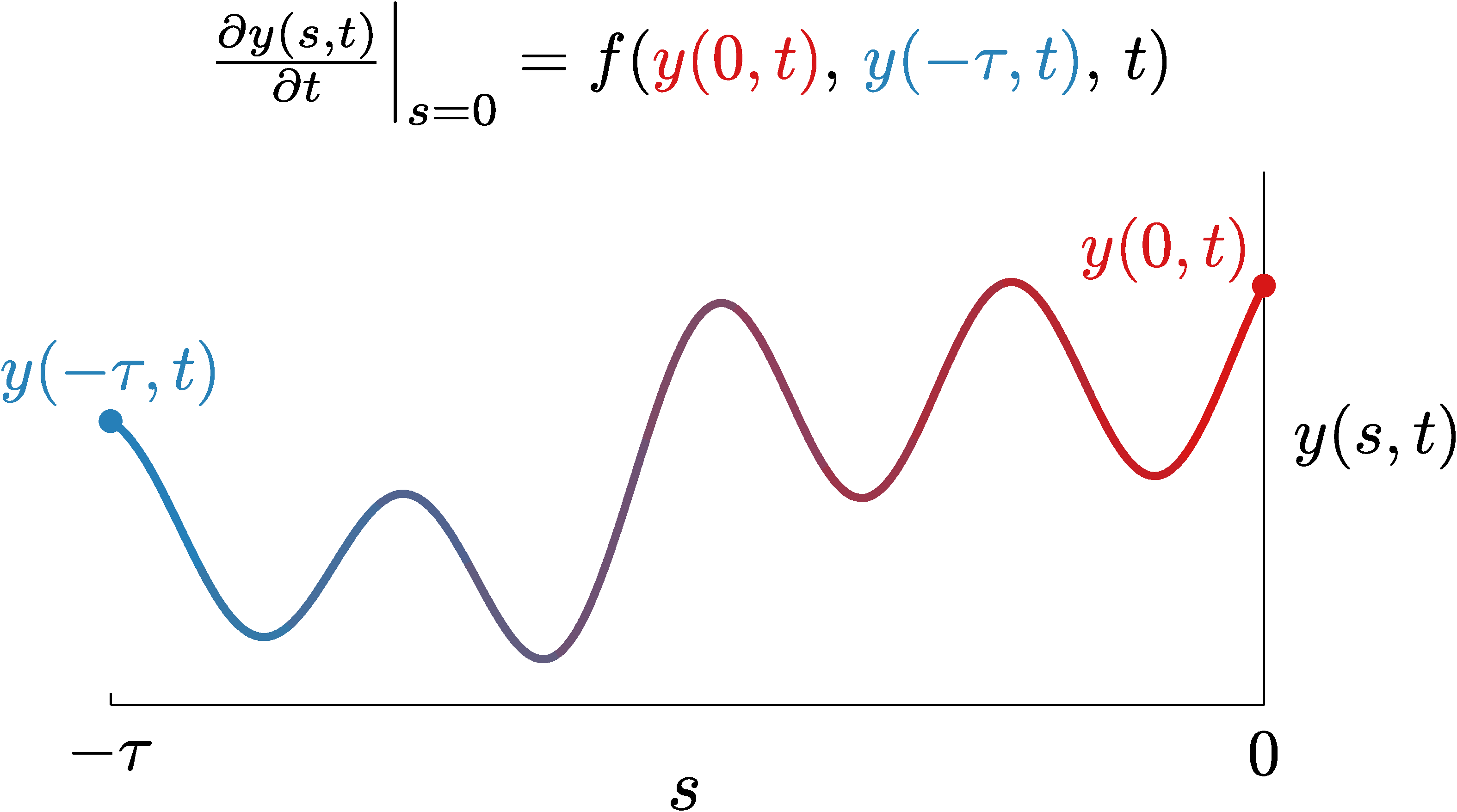}
    \end{center}
    \caption{By transforming the history function $\alpha(t), -\tau \leq t \leq 0$ into the spatial domain, we convert Eq.~\eqref{eq.dde} into a boundary control problem of the advection equation.}
    \label{fig.advection}
\end{figure}

For clarity of presentation, we shall consider a DDE with a single delay (extension of this procedure to systems of DDEs with multiple delays is trivial):
\begin{equation}
    \label{eq.dde}
    \dot{x} = f(x, x(t-\tau), t)
\end{equation}
where $\tau$ is the time delay and $x(t) = \alpha(t), -\tau \leq t \leq 0$ is the history function.
We introduce the following transformation \cite{maset2003cauchy,bellen2000cauchy}:
\begin{equation}
    \label{eq.transformation}
    y(s,t) = x(t+s)
\end{equation}
The DDE (Eq.~\ref{eq.dde}) and its history function can then be recast into the following initial--boundary value problem:
\begin{subequations}\begin{flalign}
    \label{eq.pde1}
    \frac{\partial y}{\partial t} &= \frac{\partial y}{\partial s}, \quad t \geq 0,\:\: -\tau \leq s \leq 0\\
    \label{eq.pde2}
    \left.\frac{\partial y(s,t)}{\partial t}\right|_{s=0} &= f(y(0,t),\: y(-\tau,t),\: t)\\
    \label{eq.pde3}
    y(s,0) &= \alpha(s)
\end{flalign}\label{eq.pde}\end{subequations}
The equivalent PDE representation (Eq.~\ref{eq.pde}) can be interpreted as a boundary control problem of the advection equation.
As described by Eq.~\eqref{eq.pde2} and shown in Fig.~\ref{fig.advection}, the time derivative of the solution at the right boundary is a nonlinear function of the value at the right boundary ($y(0,t)$) and the value at the left boundary ($y(-\tau,t)$).
The solution $y(s,t)$ must satisfy Eq.~\eqref{eq.pde1} over its domain $-\tau \leq s \leq 0$, the right boundary $y(0,t)$ must satisfy Eq.~\eqref{eq.pde2} for all time $t$, and the solution at time $t=0$ must satisfy the initial condition given by Eq.~\eqref{eq.pde3}.
The solution $x(t)$ of the original DDE (Eq.~\ref{eq.dde}) is simply the solution of the PDE (Eq.~\ref{eq.pde}) at the right boundary---that is, $x(t) = y(0,t)$.
In Sections \ref{subsec.mol} and \ref{subsec.galerkin}, we describe how to approximate the PDE representation (Eq.~\ref{eq.pde}) as a system of ODEs using the MoL and Galerkin approaches.

\subsection{Method of Lines (MoL) Approximation}\label{subsec.mol}
The first step in the MoL approach is to discretize the spatial domain $s \in \left[ -\tau,0 \right]$ into $N$ evenly spaced nodes $s_k \triangleq -\tau \left( k-1 \right) / \left( N-1 \right), k = 1, 2, \ldots, N$.
We then introduce auxiliary variables $y_k(t) \triangleq y(s_k, t), k = 1, 2, \ldots, N$, whereupon Eq.~\eqref{eq.pde} can be rewritten as an initial-value problem in discrete form:
\begin{subequations}\begin{flalign}
    \label{eq.mol1}
    \dot{y}_1(t) &= f(y_1(t),\: y_N(t),\: t)\\
    \label{eq.mol2}
    \dot{y}_r(t) &= \frac{y_{r-1} - y_{r+1}}{2 \Delta s}, \quad r = 2, 3, \ldots, N-1\\
    \label{eq.mol3}
    \dot{y}_N(t) &= \frac{y_{N-1} - y_N}{\Delta s}\\
    \label{eq.mol4}
    y_k(0) &= \alpha(s_k), \quad k = 1, 2, \ldots, N
\end{flalign}\label{eq.mol}\end{subequations}
We use a second-order central difference to approximate the spatial derivatives of Eq.~\eqref{eq.pde1} at nodes $s_r$ for $r = 2, 3, \ldots, N-1$ (Eq.~\ref{eq.mol2}), and a first-order forward difference at node $s_N$ (Eq.~\ref{eq.mol3}).
Equations~\eqref{eq.mol} represent an $N$-dimensional ODE approximation of the original DDE (Eq.~\ref{eq.dde}); the solution is given by $x(t) = y(0,t) = y_1(t)$.
The approximation improves as $\Delta s$ decreases (i.e., as $N$ increases).

\subsection{Galerkin Approximation}\label{subsec.galerkin}
In this approach, the solution $y(s,t)$ of Eq.~\eqref{eq.pde} is assumed to be of the following form:
\begin{equation}
    \label{eq.sersol}
    y(s,t) \approx \sum_{j=1}^{N} \phi_j(s) \eta_j(t) \triangleq \boldsymbol{\phi}^{\operatorname{T}}(s) \boldsymbol{\eta}(t)
\end{equation}
where $\boldsymbol{\phi}(s) \triangleq \big[ \phi_1(s), \phi_2(s), \ldots, \phi_N(s) \big]^{\operatorname{T}}$ is the vector of basis functions and $\boldsymbol{\eta}(t) \triangleq \big[ \eta_1(t), \allowbreak \eta_2(t), \ldots, \eta_N(t) \big]^{\operatorname{T}}$ is the vector of generalized coordinates.
We use shifted Legendre polynomials as the basis functions:
\begin{subequations}\begin{flalign}
    \label{eq.legendre1}
    \phi_1(s) &= 1\\
    \label{eq.legendre2}
    \phi_2(s) &= 1 + \frac{2s}{\tau}\\
    \label{eq.legendre3}
    \phi_j(s) &= \frac{\left( 2j-3 \right) \phi_2(s) \phi_{j-1}(s) - \left( j-2 \right) \phi_{j-2}(s)}{j-1}, \quad j = 3, 4, \ldots, N
\end{flalign}\label{eq.legendre}\end{subequations}
Substituting the series solution (Eq.~\ref{eq.sersol}) into Eq.~\eqref{eq.pde1}, pre-multiplying the result by $\boldsymbol{\phi}(s)$, and then integrating over the domain $s \in \left[ -\tau, 0 \right]$, we obtain the following:
\begin{equation}
    \label{eq.galerkin1}
    \mathbf{A} \dot{\boldsymbol{\eta}}(t) = \mathbf{B} \boldsymbol{\eta}(t)
\end{equation}
where $\mathbf{A} \triangleq \int_{-\tau}^{0} \boldsymbol{\phi}(s) \boldsymbol{\phi}^{\operatorname{T}}(s) \operatorname{d}s$ and $\mathbf{B} \triangleq \int_{-\tau}^{0} \boldsymbol{\phi}(s) \frac{\partial}{\partial s} \boldsymbol{\phi}^{\operatorname{T}}(s) \operatorname{d}s$.
Using shifted Legendre polynomials as the basis functions allows us to write the entries of matrices $\mathbf{A}$ and $\mathbf{B}$ in closed form:
\begin{subequations}\begin{flalign}
    \label{eq.matelements1}
    A_{a,b} &= \frac{\tau}{2a-1} \delta_{a,b}, \quad a,b = 1, 2, \ldots, N\\
    \label{eq.matelements2}
    B_{a,b} &= \begin{cases}
                 0, & \text{if~} a \geq b\\
                 2, & \text{if~} a+b \text{~is odd}\\
                 0, & \text{otherwise}
               \end{cases}, \quad a,b = 1, 2, \ldots, N
\end{flalign}\label{eq.matelements}\end{subequations}
We now substitute the series solution (Eq.~\ref{eq.sersol}) into the boundary condition (Eq.~\ref{eq.pde2}):
\begin{equation}
    \label{eq.galerkin2}
    \boldsymbol{\phi}^{\operatorname{T}}(0) \dot{\boldsymbol{\eta}}(t) = f( \boldsymbol{\phi}^{\operatorname{T}}(0) \boldsymbol{\eta}(t),\: \boldsymbol{\phi}^{\operatorname{T}}(-\tau) \boldsymbol{\eta}(t),\: t)
\end{equation}
The tau method is used to impose the boundary conditions while solving the ODEs.
Specifically, we replace the last row of Eq.~\eqref{eq.galerkin1} with the transformed boundary condition (Eq.~\ref{eq.galerkin2}), whereupon we obtain the following ODEs:
\begin{equation}
    \label{eq.galerkin3}
    \mathbf{A}_{\mathrm{Tau}} \dot{\boldsymbol{\eta}}(t) = \mathbf{B}_{\mathrm{Tau}} \boldsymbol{\eta}(t) + \mathbf{f}_{\mathrm{Tau}}(t)
\end{equation}
where $\mathbf{A}_{\mathrm{Tau}}$, $\mathbf{B}_{\mathrm{Tau}}$, and $\mathbf{f}_{\mathrm{Tau}}$ are defined as follows:
\begin{equation}
    \label{eq.galerkin4}
    \mathbf{A}_{\mathrm{Tau}} \triangleq \begin{bmatrix}
                                           \hat{\mathbf{A}}\\
                                           \boldsymbol{\phi}^{\operatorname{T}}(0)
                                         \end{bmatrix}, \quad
    \mathbf{B}_{\mathrm{Tau}} \triangleq \begin{bmatrix}
                                           \hat{\mathbf{B}}\\
                                           0
                                         \end{bmatrix}, \quad
    \mathbf{f}_{\mathrm{Tau}} \triangleq \begin{Bmatrix}
                                           \mathbf{0}\\
                                           f( \boldsymbol{\phi}^{\operatorname{T}}(0) \boldsymbol{\eta}(t),\: \boldsymbol{\phi}^{\operatorname{T}}(-\tau) \boldsymbol{\eta}(t),\: t)
                                         \end{Bmatrix}
\end{equation}
Matrices $\hat{\mathbf{A}}$ and $\hat{\mathbf{B}}$ in Eq.~\eqref{eq.galerkin4} are obtained by deleting the last row of $\mathbf{A}$ and $\mathbf{B}$, respectively.
We calculate the initial conditions $\boldsymbol{\eta}(0)$ for solving Eq.~\eqref{eq.galerkin3} by substituting the series solution (Eq.~\ref{eq.sersol}) into Eq.~\eqref{eq.transformation}.
We then pre-multiply by $\boldsymbol{\phi}(s)$ and integrate over the domain $s \in \left[ -\tau, 0 \right]$:
\begin{equation}
    \label{eq.galerkin5}
    \boldsymbol{\eta}(t) = \mathbf{A}^{-1} \int_{-\tau}^{0} \boldsymbol{\phi}(s) x(t+s) \operatorname{d}s
\end{equation}
Finally, we substitute $t=0$ into Eq.~\eqref{eq.galerkin5} to obtain the new initial conditions:
\begin{equation}
    \label{eq.galerkin6}
    \boldsymbol{\eta}(0) = \mathbf{A}^{-1} \int_{-\tau}^{0} \boldsymbol{\phi}(s) \alpha(s) \operatorname{d}s
\end{equation}
We can now solve Eq.~\eqref{eq.galerkin3} for $\boldsymbol{\eta}(t)$ using the initial conditions given by Eq.~\eqref{eq.galerkin6}; the solution of the original DDE is given by $x(t) = y(0,t) = \boldsymbol{\phi}^{\operatorname{T}}(0) \boldsymbol{\eta}(t)$.
To summarize, we have converted the nonlinear DDE (Eq.~\ref{eq.dde}) and its history function $x(t) = \alpha(t), -\tau \leq t \leq 0$ into a system of first-order ODEs (Eq.~\ref{eq.galerkin3}) with the initial conditions given by Eq.~\eqref{eq.galerkin6}.

\subsection{Computing Lyapunov Exponents}\label{subsec.lyapunov}
In Sections \ref{subsec.mol} and \ref{subsec.galerkin}, we formed ODE approximations of DDEs.
In this section, we describe a technique for computing the Lyapunov exponents $\lambda_i, i = 1, 2, \ldots, N$ for $N$-dimen\-sional systems of ODEs \cite{wolf1985lyapunov,sandri1996lyapunov}:
\begin{equation}
    \label{eq.ode}
    \dot{\boldsymbol{\eta}}(t) = \mathbf{h}(\boldsymbol{\eta}(t),\: t)
\end{equation}
where $\mathbf{h}(\boldsymbol{\eta}(t),\: t) \triangleq \mathbf{A}_{\mathrm{Tau}}^{-1} \left( \mathbf{B}_{\mathrm{Tau}} \boldsymbol{\eta}(t) + \mathbf{f}_{\mathrm{Tau}}(t) \right)$ from the Galerkin approximation method (Eq.~\ref{eq.galerkin3}).
For these ODEs (Eq.~\ref{eq.ode}), the evolution of an initial infinitesimal perturbation from a reference trajectory satisfies the following matrix differential equation:
\begin{equation}
    \label{eq.perturb}
    \dot{\boldsymbol{\Psi}}(t) = \operatorname{J}_{\eta} \mathbf{h}(\boldsymbol{\eta}(t),\: t) \cdot \boldsymbol{\Psi}(t), \quad \boldsymbol{\Psi}(0) = \mathbf{I}
\end{equation}
where $\operatorname{J}_{\boldsymbol{\eta}} \mathbf{h}(\boldsymbol{\eta}(t),\: t)$ is the Jacobian of $\mathbf{h}(\boldsymbol{\eta}(t),\: t)$ with respect to $\boldsymbol{\eta}$ and $\mathbf{I}$ is the identity matrix.
Equation~\eqref{eq.perturb} represents the sensitivity of Eq.~\eqref{eq.ode} with respect to the initial conditions \cite{nayfeh1995applied} and is unidirectionally coupled to Eq.~\eqref{eq.ode}; these equations must be integrated simultaneously.

An iterative method \cite{wolf1985lyapunov,sandri1996lyapunov} is used to calculate the Lyapunov exponents of Eq.~\eqref{eq.ode}; we describe this method here for completeness.
A sequence of $K$ integrations is performed on a system comprised of Eqs.~\eqref{eq.ode} and \eqref{eq.perturb}; each integration is of duration $T$.
In the first iteration, we integrate from $t=0$ to $t=T$ to arrive at $\boldsymbol{\eta}(T)$ and $\boldsymbol{\Psi}(T)$.
In the second iteration, we integrate from $t=T$ to $t=2T$ using the final state from the first iteration $\boldsymbol{\eta}(T)$ as the initial condition for Eq.~\eqref{eq.ode}.
The initial condition for Eq.~\eqref{eq.perturb} is obtained by performing Gram--Schmidt orthonormalization on $\boldsymbol{\Psi}(T)$, which makes this iterative method stable.
This procedure is continued for $K$ iterations, whereupon the $N$ Lyapunov exponents can be calculated as follows:
\begin{equation}
    \label{eq.lyapunov}
    \lambda_j = \frac{1}{KT} \sum_{k=1}^{K} \log \left( \left\Vert \psi_j(kT) \right\Vert \right), \quad j = 1, 2, \ldots, N
\end{equation}
where $\psi_j(kT)$ is the $j$th column of matrix $\boldsymbol{\Psi}(kT)$.
For sufficiently large $K$, the $\lambda_j$ converge to the Lyapunov exponents of the system.

\section{Results and Discussion}\label{sec.results}
In this section, we present numerical results using the theory described in Section \ref{sec.math}.
We compute Lyapunov exponents and strange attractors of nonlinear DDEs using the standard MoL approach and the proposed approach using Galerkin projection.
We use the iterative procedure described in Section \ref{subsec.lyapunov} with duration $T=0.5$ to compute Lyapunov exponents.
Note that chaotic responses obtained from direct numerical integration of DDEs will always differ from the responses of approximating ODE systems.
This discrepancy is unavoidable because chaotic responses are sensitive to small differences between mathematical models and initial conditions.
All results were generated in MATLAB using the \textit{ode15s}, \textit{dde23}, and \textit{rk4} integrators with equal relative and absolute integration tolerances (``tol.''), as noted below.

\subsection{First-order, Nonlinear DDE}\label{subsec.ex1}
\begin{figure}[t]
    \begin{center}
        \includegraphics[width=0.9\textwidth]{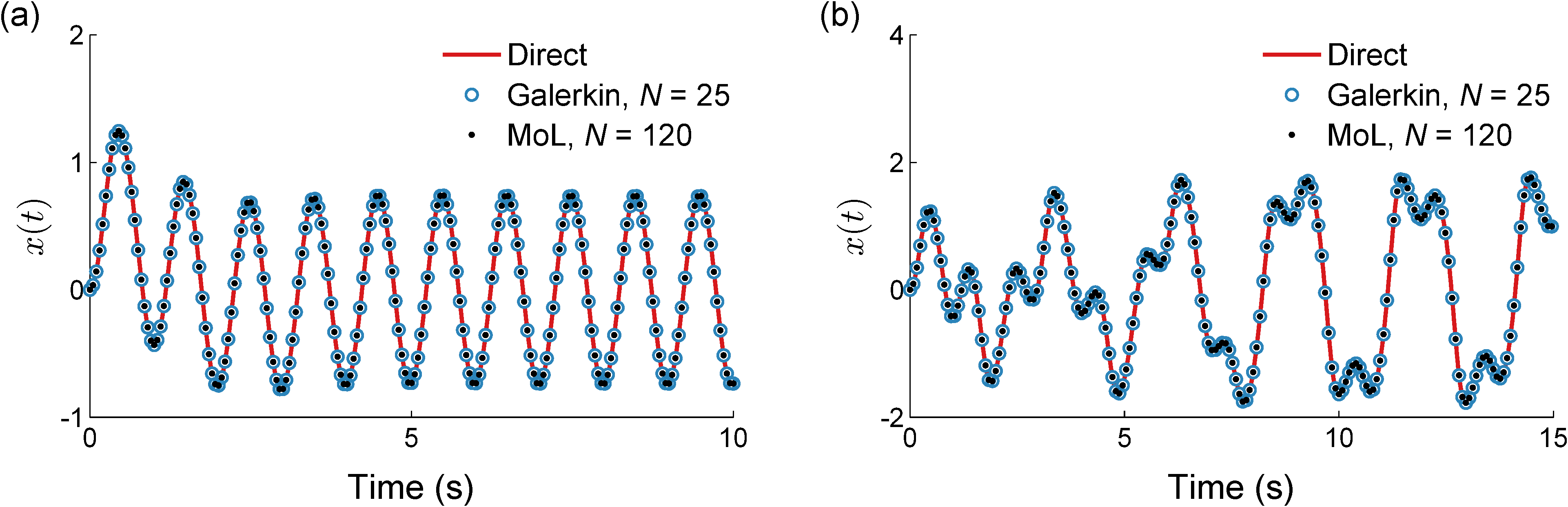}
    \end{center}
    \caption{Time response of Eq.~\eqref{eq.ex1_dde} obtained using the \textit{dde23} solver (``Direct''), and from the Galerkin and MoL approximations using \textit{ode15s} (all tols.\ $10^{-9}$). Simulations were performed for two values of parameter $a$: (a) $0.4968$ and (b) $2.7705$.}
    \label{fig.ex1_response}
\end{figure}

\begin{figure}[t]
    \begin{center}
        \includegraphics[width=0.4\textwidth]{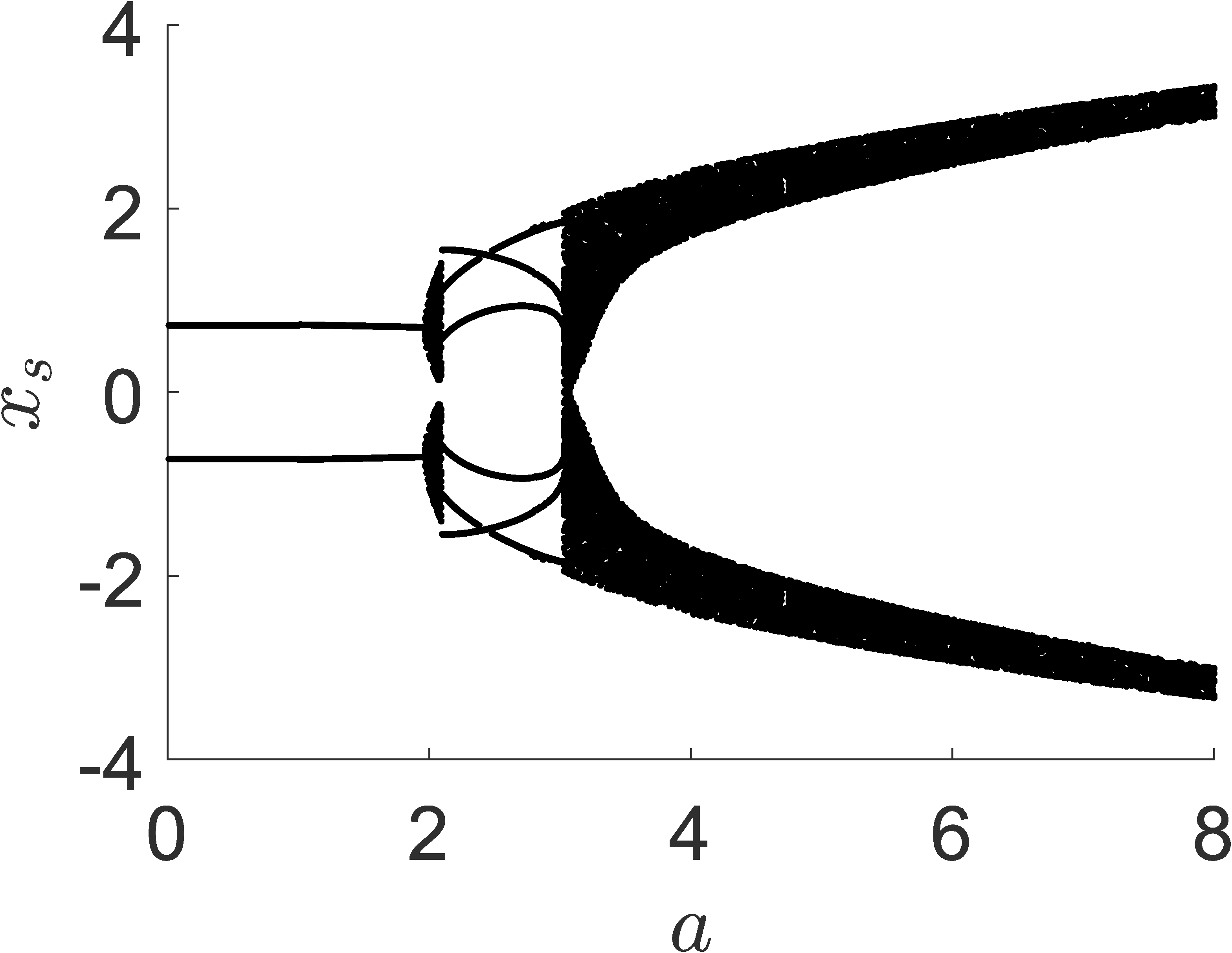}
    \end{center}
    \caption{Bifurcation diagram for the first-order, nonlinear system described by Eq.~\eqref{eq.ex1_dde}. The steady-state displacement $x_s$ is shown as parameter $a$ varies. The plot was generated using \textit{dde23} (tol.\ $10^{-6}$).}
    \label{fig.ex1_bifurcation}
\end{figure}

\begin{figure}[t]
    \begin{center}
        \includegraphics[width=0.45\textwidth]{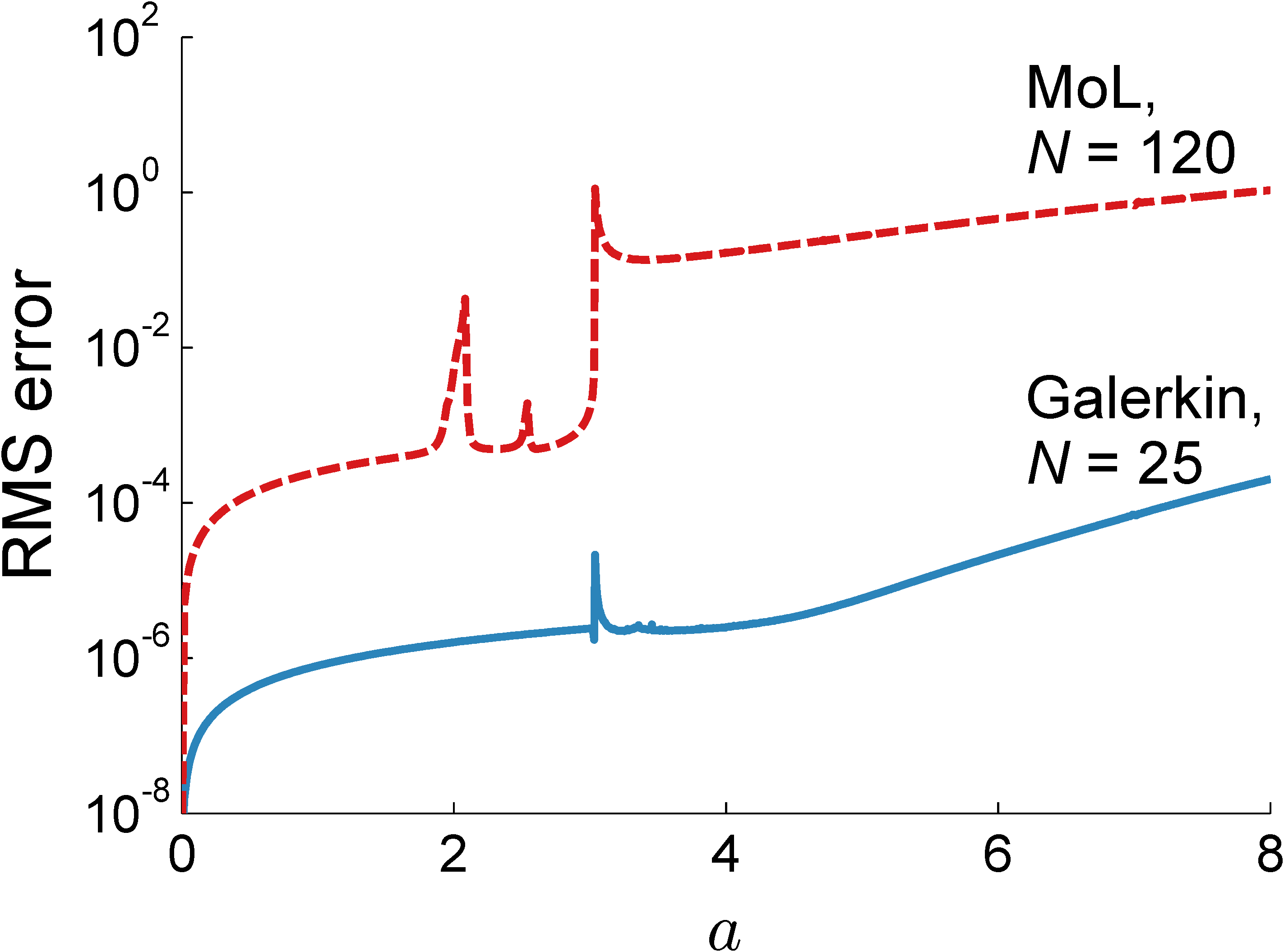}
    \end{center}
    \caption{Root-mean-square (RMS) error of MoL and Galerkin approximations using \textit{ode15s} relative to a direct solution of Eq.~\eqref{eq.ex1_dde} using \textit{dde23} as parameter $a$ varies (all tols.\ $10^{-9}$). The Galerkin approach resulted in substantially lower RMS error with a system of much lower dimension.}
    \label{fig.ex1_error}
\end{figure}

We first consider the following first-order, nonlinear DDE \cite{wahi2005vibrations}:
\begin{equation}
    \label{eq.ex1_dde}
    \dot{x}(t) = -ax(t-\tau) - bx(t)^3 + f_0 \sin(\omega t)
\end{equation}
where $\tau=1$, $b=\pi/4$, $f_0=3\pi/2$, and $\omega=2\pi$.
To compare the Galerkin and MoL approaches, we plot the time responses $x(t)$ for two values of $a$ in Fig.~\ref{fig.ex1_response}.
As shown, the Galerkin approximation with $N=25$ terms in the series solution shows good agreement with the direct solution as well as the MoL approximation.
The bifurcation diagram for this system is shown in Fig.~\ref{fig.ex1_bifurcation}.
Depending on the value of parameter $a$, the solution may be periodic, quasiperiodic, or chaotic.
Regardless, the error in the Galerkin solution is low for all values of $a$, which we demonstrate by computing the root-mean-square (RMS) error between direct solutions $x(t)$ and approximate solutions $\hat{x}(t)$:
\begin{equation}
    \label{eq.ex1_rmse}
    \text{RMS~error} \triangleq \sqrt{ \frac{\sum_{i=1}^{n} \left( x(t_i) - \hat{x}(t_i) \right)^2}{n} }
\end{equation}
where $n$ is the number of time points in each solution.
We evaluate Eq.~\eqref{eq.ex1_rmse} using 500-second simulations and $n=5 \times 10^4$ equally spaced time points.
In Fig.~\ref{fig.ex1_error}, we illustrate a key benefit of the Galerkin approach: the RMS error relative to the direct solution is substantially lower when using the Galerkin method with $N=25$ than when using the MoL method with $N=120$.

\subsection{Mackey--Glass Equation}\label{subsec.ex2}
\begin{figure}[t]
    \begin{center}
        \includegraphics[width=0.9\textwidth]{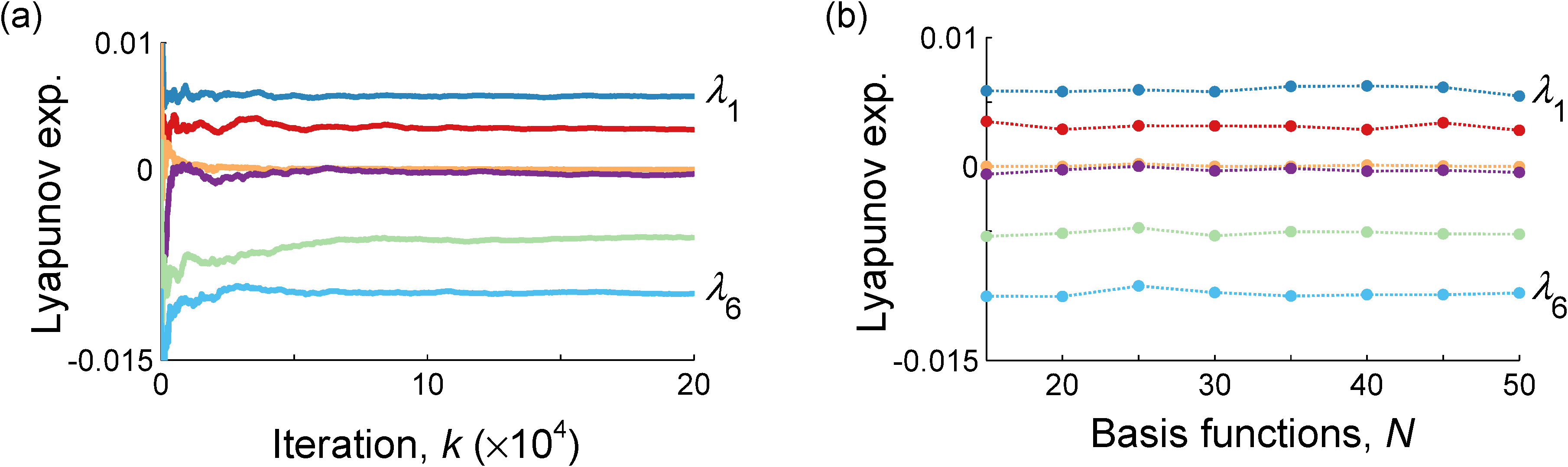}
    \end{center}
    \caption{Dominant Lyapunov exponents of the Mackey--Glass equation (Eq.~\ref{eq.ex2_dde}) computed using Galerkin approximation with \textit{ode15s} (tol.\ $10^{-6}$). The values of the Lyapunov exponents are shown (a) as $k$ increases for $N=30$ and (b) as $N$ increases.}
    \label{fig.ex2_exponents}
\end{figure}

We now consider the Mackey--Glass equation, which is used to model blood production \cite{sigeti1995decay,breda2014lyapunov}:
\begin{equation}
    \label{eq.ex2_dde}
    \dot{x}(t) = \frac{ax(t-\tau)}{1 + \left[ x(t-\tau) \right]^c} - bx(t)
\end{equation}
where $a=0.2$, $b=0.1$, $c=10$, and $\tau=50$.
In this example, we apply the Galerkin approximation and compute the six dominant Lyapunov exponents.
Shown in Fig.~\ref{fig.ex2_exponents}(a) are these Lyapunov exponents converging over $k$ for a particular value of $N$ (see Section \ref{subsec.lyapunov}); Fig.~\ref{fig.ex2_exponents}(b) shows the converged values of the Lyapunov exponents with respect to the number of terms in the series solution $N$.
As shown in Table \ref{tbl.ex2}, the Lyapunov exponents computed using the Galerkin method compare favorably with those reported in the literature \cite{sigeti1995decay,breda2014lyapunov}.
\begin{table}[t]
    \centering
    \caption{Dominant Lyapunov exponents of the Mackey--Glass equation (Eq.~\ref{eq.ex2_dde}) computed using Galerkin approximation and as reported by Breda and Van Vleck \cite{breda2014lyapunov} and Sigeti \cite{sigeti1995decay}.}
    \begin{tabular}{lllll}
        \hline\noalign{\smallskip}
        & \multicolumn{2}{l}{Galerkin ($\times 10^{-3}$)} & Breda, Table 2 & Sigeti, Table 1\\
        & $N=20$ & $N=50$ & ($\times 10^{-3}$) & ($\times 10^{-3}$)\\
        \noalign{\smallskip}\hline\noalign{\smallskip}
        $\lambda_1$ & $\phantom{+1}5.811$ & $\phantom{+}5.456$ & $\phantom{+}5.76$ & $\phantom{+}5.83$\\
        $\lambda_2$ & $\phantom{+1}2.890$ & $\phantom{+}2.802$ & $\phantom{+}3.02$ & $\phantom{+}3.15$\\
        $\lambda_3$ & $\phantom{+1}0.007$ & $\phantom{+}0.002$ & $\phantom{+}0.65$ & $\phantom{+}0.01$\\
        $\lambda_4$ & \setlength{\dimen0}{\widthof{$1$}}\hspace{\dimen0}$-0.247$ & $-0.447$ & $-0.85$ & $-0.29$\\
        $\lambda_5$ & \setlength{\dimen0}{\widthof{$1$}}\hspace{\dimen0}$-5.159$ & $-5.229$ & $-4.78$ & $-5.08$\\
        $\lambda_6$ & $-10.058$ & $-9.784$ & $-9.85$ & $-9.78$\\
        \noalign{\smallskip}\hline
    \end{tabular}
    \label{tbl.ex2}
\end{table}

\subsection{``Nearly-Brownian'' Chaotic System}\label{subsec.ex3}
\begin{figure}[t]
    \begin{center}
        \includegraphics[width=0.5\textwidth]{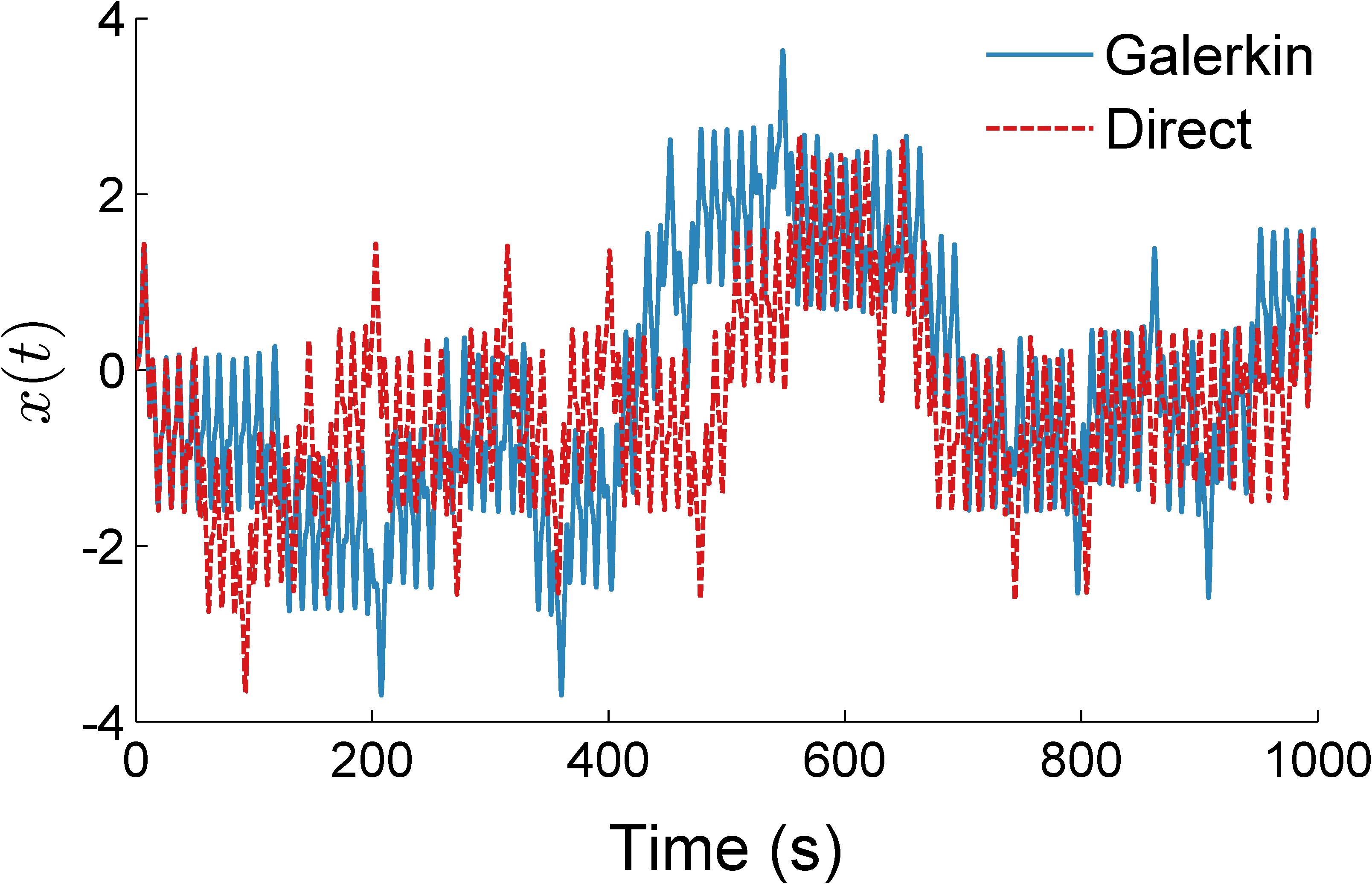}
    \end{center}
    \caption{Time response of the ``nearly-Brownian'' chaotic system (Eq.~\ref{eq.ex3_dde}) obtained using a fourth-order Runge--Kutta method (``Direct'') with 10-millisecond time steps, and using the Galerkin approximation with \textit{ode15s} (tol.\ $10^{-6}$). The history function was approximated in the direct method using spline interpolation.}
    \label{fig.ex3_response}
\end{figure}

\begin{figure}[t]
    \begin{center}
        \includegraphics[width=0.85\textwidth]{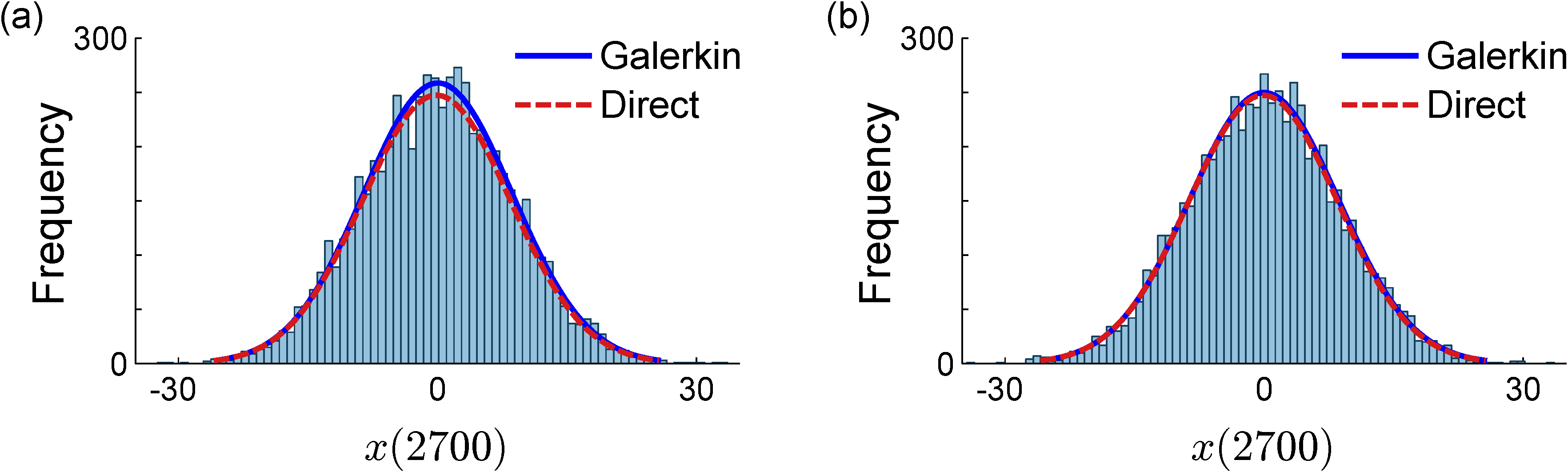}
    \end{center}
    \caption{Histograms of the final state $x(2700)$ of the ``nearly-Brownian'' chaotic system (Eq.~\ref{eq.ex3_dde}) obtained using the Galerkin approach with (a) $N=10$ and (b) $N=20$ (\textit{ode15s}; tol.\ $10^{-6}$). Also shown are normal density functions fit to the histograms obtained using the Galerkin and direct solutions.}
    \label{fig.ex3_histogram}
\end{figure}

\begin{figure}[t]
    \begin{center}
        \includegraphics[width=0.9\textwidth]{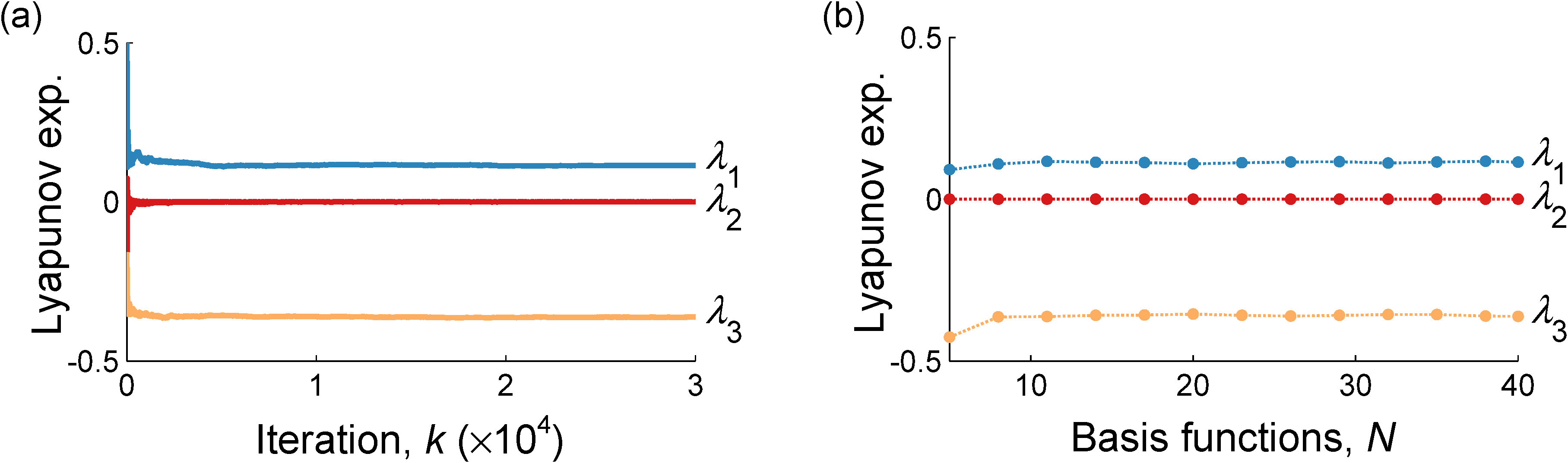}
    \end{center}
    \caption{Dominant Lyapunov exponents of the ``nearly-Brownian'' chaotic system (Eq.~\ref{eq.ex3_dde}) computed using Galerkin approximation (\textit{ode15s}; tol.\ $10^{-6}$). The values of the Lyapunov exponents are shown (a) as $k$ increases for $N=30$ and (b) as $N$ increases.}
    \label{fig.ex3_lyapunov}
\end{figure}

We now consider the ``nearly-Brownian'' chaotic system \cite{chekroun2016galerkin}:
\begin{equation}
    \label{eq.ex3_dde}
    \dot{x}(t) = a \sin \left( \int_{t-\tau}^{t} x(s) \operatorname{d}s \right)
\end{equation}
where $a=0.5$ and $\tau=5.5$.
As shown in Fig.~\ref{fig.ex3_response}, the response $x(t)$ obtained using Galerkin approximation matches the direct solution, but for only the first 50 seconds; the solutions then diverge due to the chaotic nature of the system.
We, therefore, use statistical methods to compare the Galerkin and direct solutions.
We performed a large number of simulations (6300) using the Galerkin and direct approaches, with history functions chosen randomly from the range $\left[ -1, 1 \right]$, and compared the final state $x(t)$ at $t=2700$.
As shown in Fig.~\ref{fig.ex3_histogram}, the Galerkin approximation preserves the statistical properties of the solution and converges with increasing $N$.
We also evaluate the three dominant Lyapunov exponents of Eq.~\eqref{eq.ex3_dde} using the proposed Galerkin method.
Once again, the Lyapunov exponents converge over iteration number $k$ for a particular value of $N$ (Fig.~\ref{fig.ex3_lyapunov}(a)), and the converged Lyapunov exponents remain similar for sufficiently large $N$ (Fig.~\ref{fig.ex3_lyapunov}(b)).
With $N=40$ terms in the series solution, the computed Lyapunov exponents are $\lambda_1 = 0.114$, $\lambda_2 = 0$, and $\lambda_3 = -0.362$.
Note that $\lambda_1 > 0$, which correctly indicates that the system is chaotic.

\subsection{Bimodal Chaotic System}\label{subsec.ex4}
\begin{figure}[t]
    \begin{center}
        \includegraphics[width=0.5\textwidth]{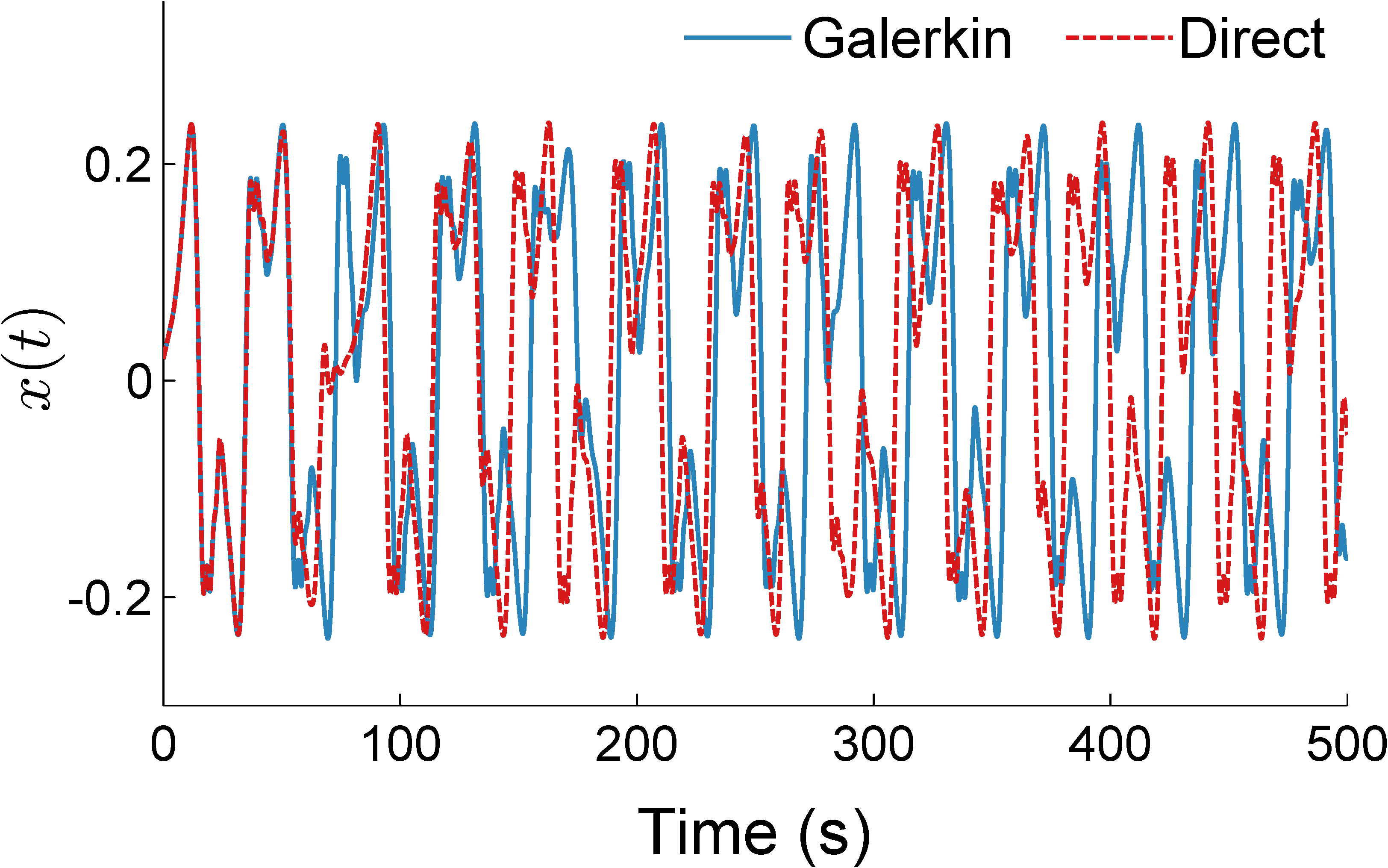}
    \end{center}
    \caption{Time response of the bimodal chaotic system (Eq.~\ref{eq.ex4_dde}) obtained using the \textit{dde23} solver (``Direct'') and the Galerkin approximation with \textit{ode15s} (all tols.\ $10^{-6}$).}
    \label{fig.ex4_response}
\end{figure}

\begin{figure}[t]
    \begin{center}
        \includegraphics[width=0.7\textwidth]{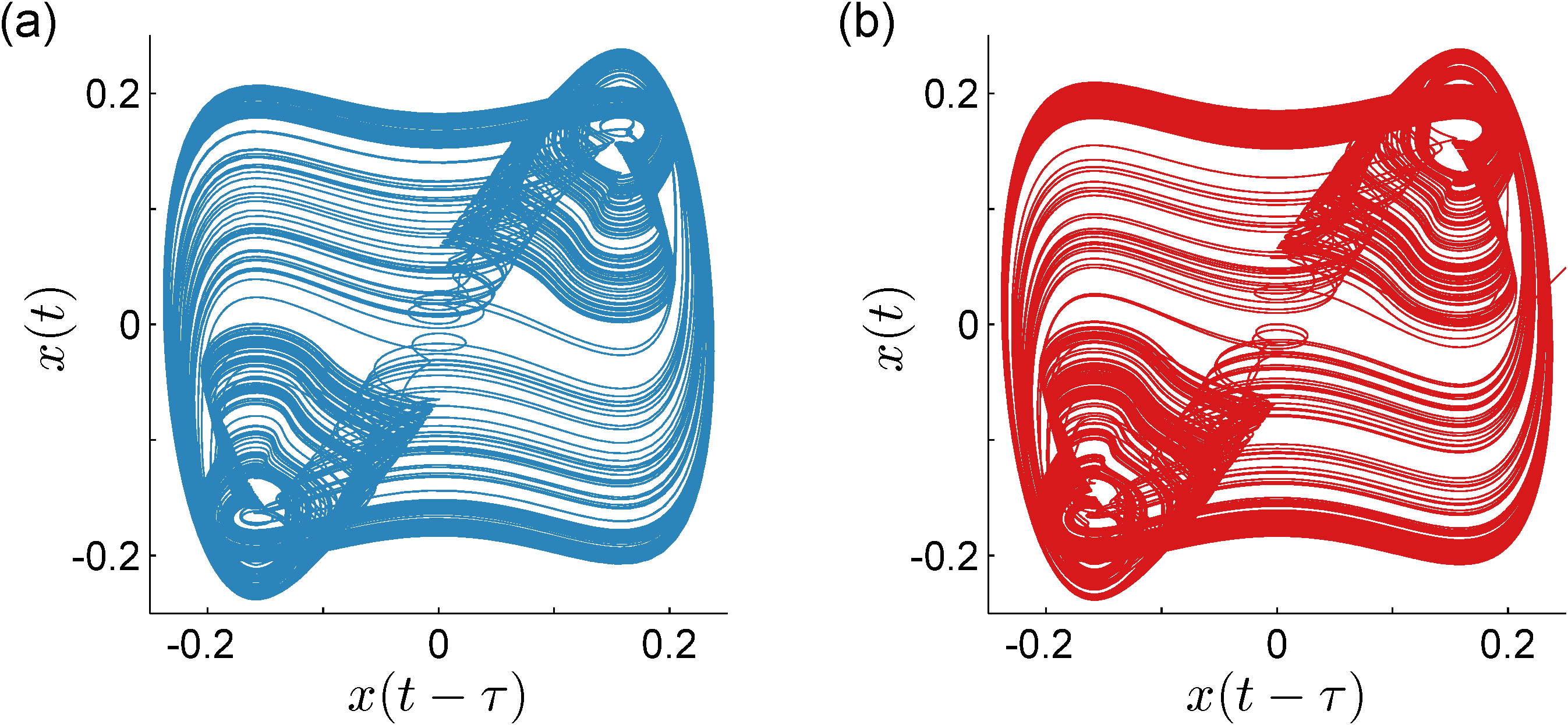}
    \end{center}
    \caption{Strange attractor of the bimodal chaotic system (Eq.~\ref{eq.ex4_dde}) obtained using (a) Galerkin approximation with \textit{ode15s} and (b) the \textit{dde23} solver (all tols.\ $10^{-6}$).}
    \label{fig.ex4_attractor}
\end{figure}

\begin{figure}[t]
    \begin{center}
        \includegraphics[width=0.9\textwidth]{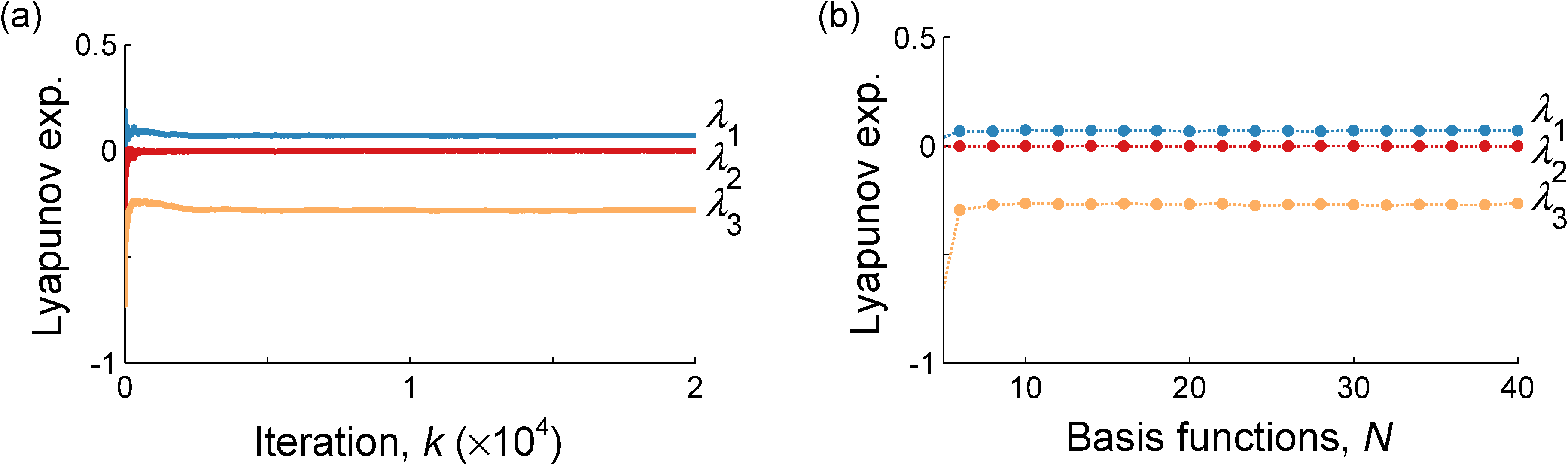}
    \end{center}
    \caption{Dominant Lyapunov exponents of the bimodal chaotic system (Eq.~\ref{eq.ex4_dde}) computed using Galerkin approximation with \textit{ode15s} (tol.\ $10^{-6}$). The values of the Lyapunov exponents are shown (a) as $k$ increases for $N=40$ and (b) as $N$ increases.}
    \label{fig.ex4_lyapunov}
\end{figure}

In this example, we consider the following bimodal chaotic system, which includes a time-delayed term with cubic nonlinearity \cite{chekroun2016galerkin}:
\begin{equation}
    \label{eq.ex4_dde}
    \dot{x}(t) = a x(t-\tau) - b x(t-\tau)^3
\end{equation}
where $a=0.5$, $b=20$, and $\tau=3.35$.
As shown in Fig.~\ref{fig.ex4_response}, the response $x(t)$ obtained using Galerkin approximation matches the direct solution at the beginning of the simulation; however, as was the case in the previous example, the solutions eventually diverge due to the chaotic nature of the system.
In Fig.~\ref{fig.ex4_attractor}, the strange attractor of Eq.~\eqref{eq.ex4_dde} is shown from the Galerkin and direct solutions in the phase space of $x(t)$ and $x(t-\tau)$; the attractors are similar in both extent and overall appearance.
As shown in Fig.~\ref{fig.ex4_lyapunov}, the Galerkin method converges to the following dominant Lyapunov exponents when $N=40$ terms are used in the series solution: $\lambda_1 = 0.071$, $\lambda_2 = 0$, and $\lambda_3 = -0.264$.
Again, we note that the dominant Lyapunov exponent is positive, which indicates that the system is chaotic.

\subsection{El Ni{\~n}o--Southern Oscillation (ENSO) Model}\label{subsec.ex5}
\begin{figure}[t]
    \begin{center}
        \includegraphics[width=0.7\textwidth]{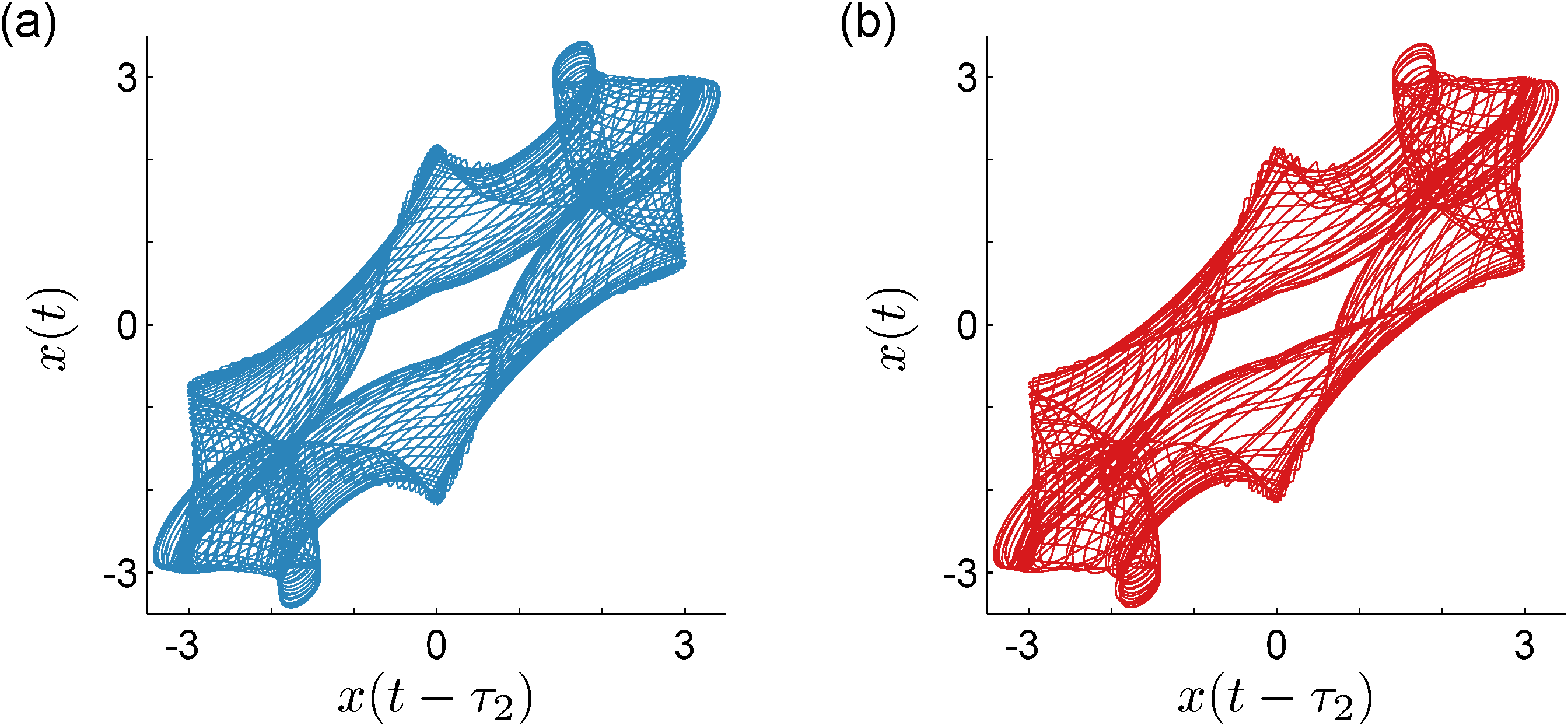}
    \end{center}
    \caption{Strange attractor of El Ni{\~n}o--Southern Oscillation model (Eq.~\ref{eq.ex5_dde}) obtained using (a) Galerkin approximation with \textit{ode15s} and (b) the \textit{dde23} solver (all tols.\ $10^{-6}$).}
    \label{fig.ex5_attractor}
\end{figure}

\begin{figure}[t]
    \begin{center}
        \includegraphics[width=0.9\textwidth]{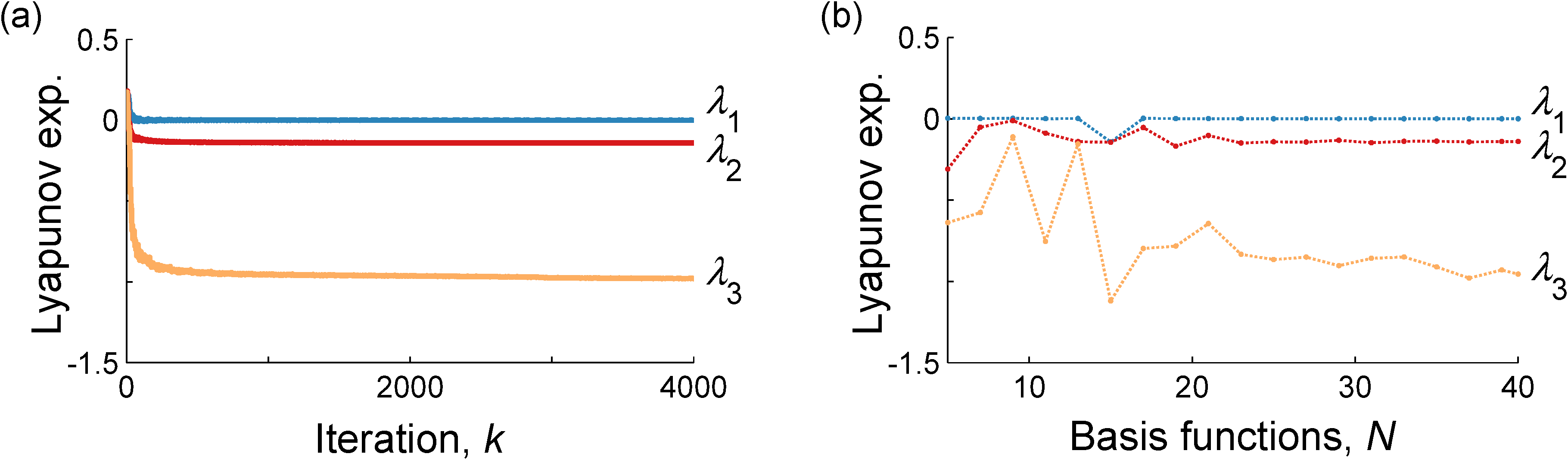}
    \end{center}
    \caption{Dominant Lyapunov exponents of the El Ni{\~n}o--Southern Oscillation model (Eq.~\ref{eq.ex5_dde}) computed using Galerkin approximation (\textit{ode15s}; tol.\ $10^{-6}$). The values of the Lyapunov exponents are shown (a) as $k$ increases for $N=40$ and (b) as $N$ increases.}
    \label{fig.ex5_lyapunov}
\end{figure}

\begin{figure}[t]
    \begin{center}
        \includegraphics[width=0.5\textwidth]{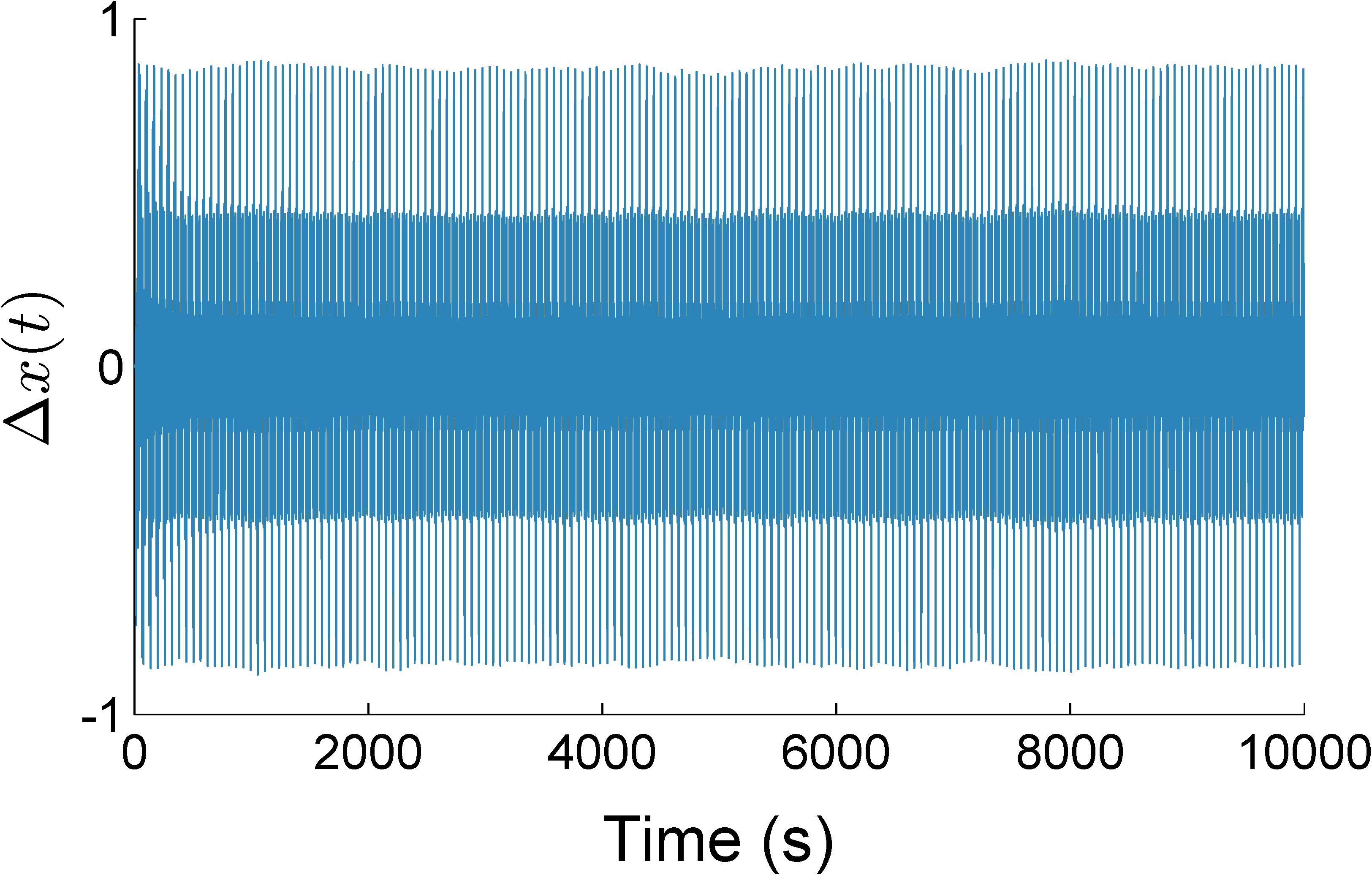}
    \end{center}
    \caption{Difference between the time responses of the El Ni{\~n}o--Southern Oscillation model (Eq.~\ref{eq.ex5_dde}) using history values 0.025 and 0.025001. The simulations were performed using the \textit{dde23} solver (tol.\ $10^{-6}$).}
    \label{fig.ex5_diff}
\end{figure}

The ENSO phenomenon is a major source of variability in regional climate patterns.
An idealized mathematical model for ENSO is given as follows \cite{chekroun2016galerkin}:
\begin{equation}
    \label{eq.ex5_dde}
    \dot{x}(t) = -\alpha \tanh(\kappa x(t-\tau_1)) + \beta \tanh(\kappa x(t-\tau_2)) + \gamma \cos(2 \pi t)
\end{equation}
where $\alpha=2.1$, $\beta=1.05$, $\gamma=3$, $\kappa=10$, $\tau_1=0.95$, and $\tau_2=5.13$.
In Fig.~\ref{fig.ex5_attractor}, the strange attractor of Eq.~\eqref{eq.ex5_dde} is shown from the Galerkin and direct solutions in the phase space of $x(t)$ and $x(t-\tau_2)$; we again confirm that the attractors are similar in both extent and overall appearance.
The first three Lyapunov exponents converged to $\lambda_1 = 0$, $\lambda_2 = -0.140$, and $\lambda_3 = -0.955$ when $N=40$, as shown in Fig.~\ref{fig.ex5_lyapunov}.
Note that the dominant Lyapunov exponent is zero in this case, which indicates that the system is \emph{not} chaotic with the parameter values listed above.
As confirmation, we simulate Eq.~\eqref{eq.ex5_dde} using two slightly different history values: 0.025 and 0.025001; if the system were chaotic, these solutions would diverge.
As shown in Fig.~\ref{fig.ex5_diff}, the difference between these trajectories $\Delta x(t)$ is bounded, confirming that Eq.~\eqref{eq.ex5_dde} is not chaotic with these parameter values.

\subsection{Delayed Lorenz Attractor}\label{subsec.ex6}
\begin{figure}[t]
    \begin{center}
        \includegraphics[width=0.7\textwidth]{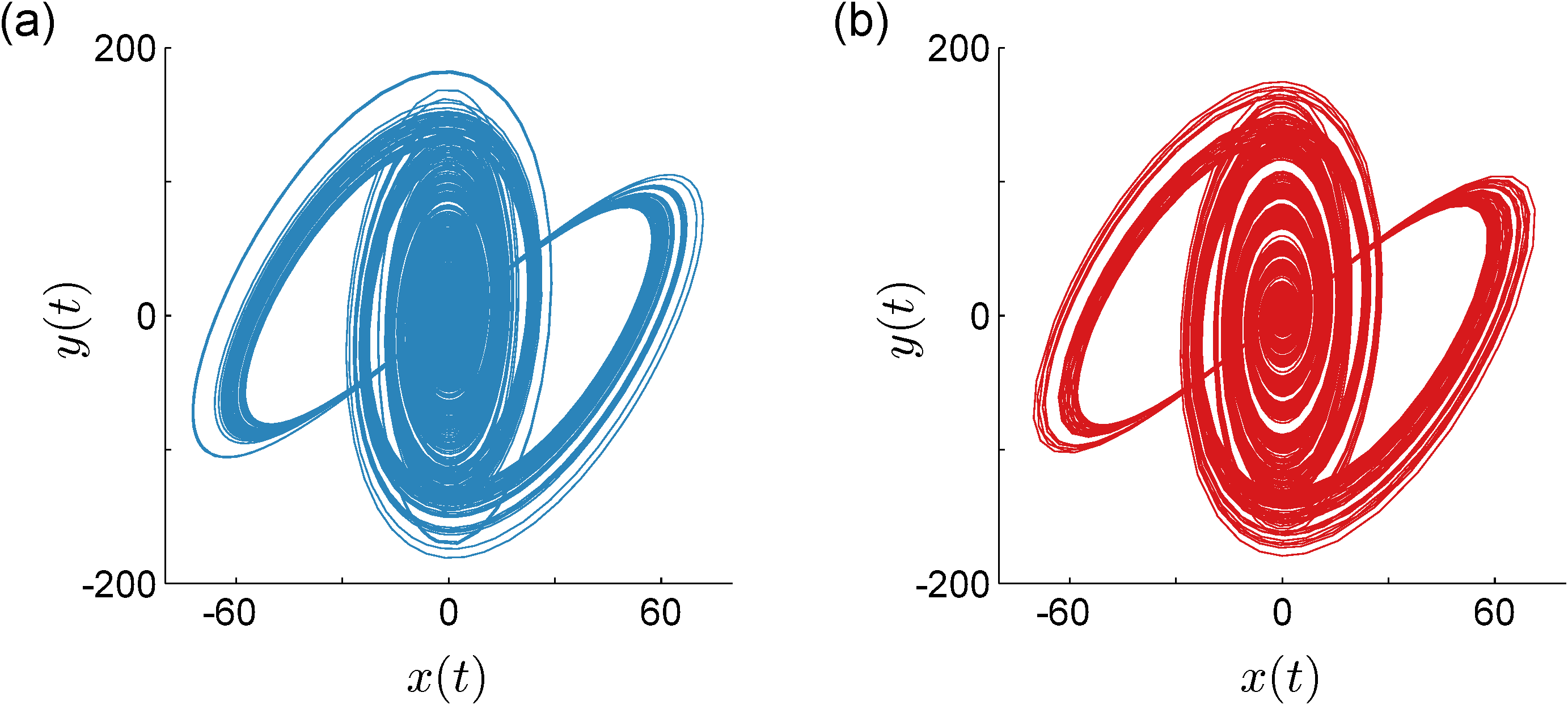}
    \end{center}
    \caption{Strange attractor of the delayed Lorenz attractor (Eq.~\ref{eq.ex6_sys}) with $\tau=0.13$, obtained using (a) the Galerkin approximation with \textit{ode15s} and (b) the \textit{dde23} solver (all tols.\ $10^{-6}$).}
    \label{fig.ex6_attractor1}
\end{figure}

\begin{figure}[t]
    \begin{center}
        \includegraphics[width=0.9\textwidth]{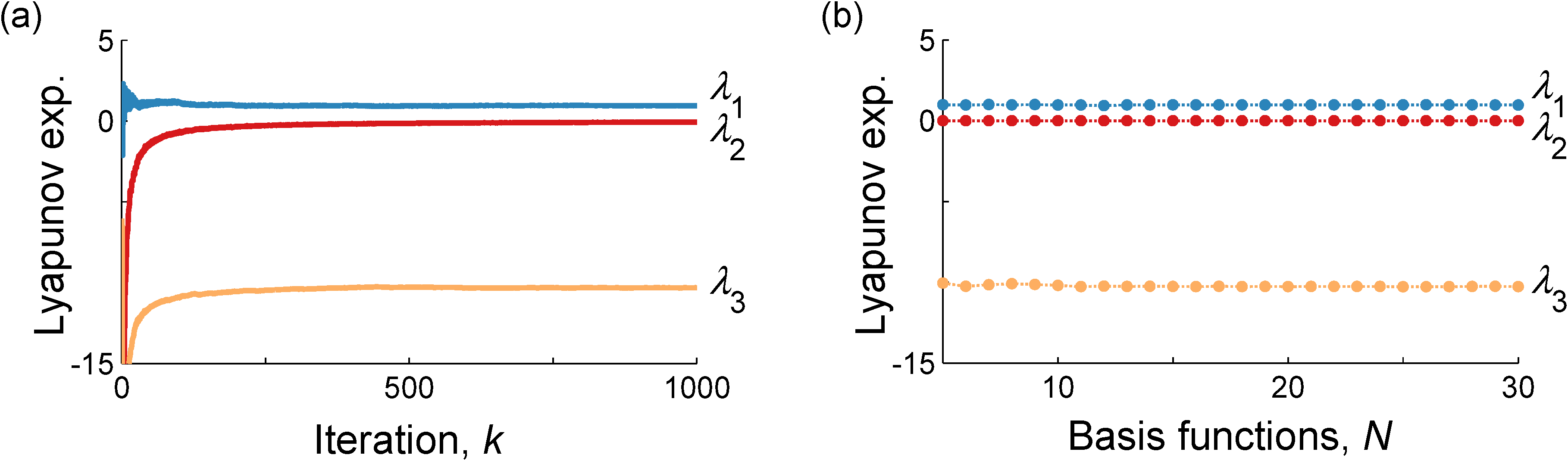}
    \end{center}
    \caption{Dominant Lyapunov exponents of the delayed Lorenz attractor (Eq.~\ref{eq.ex6_sys}) with $\tau=0.13$, computed using Galerkin approximation (\textit{ode15s}; tol.\ $10^{-6}$). The values of the Lyapunov exponents are shown (a) as $k$ increases for $N=20$ and (b) as $N$ increases.}
    \label{fig.ex6_lyapunov1}
\end{figure}

\begin{figure}[t]
    \begin{center}
        \includegraphics[width=0.7\textwidth]{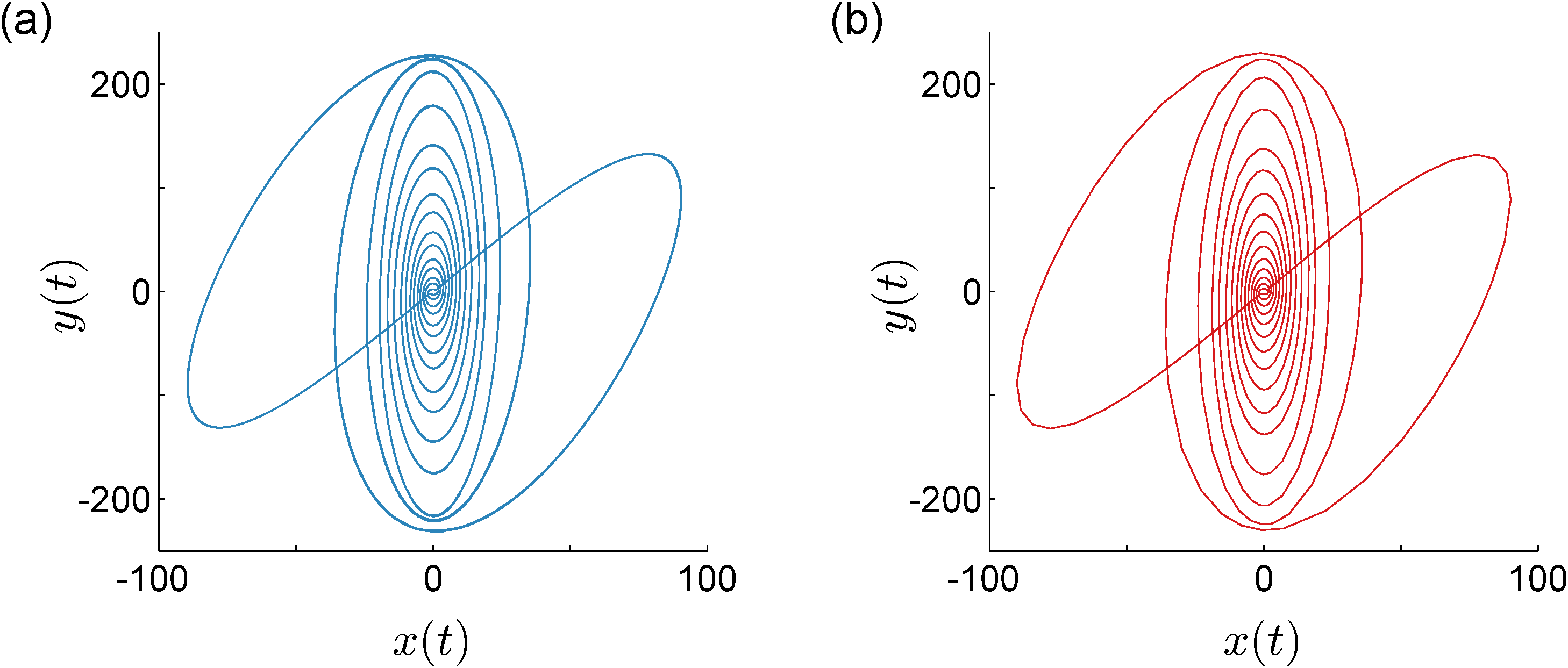}
    \end{center}
    \caption{Strange attractor of the delayed Lorenz attractor (Eq.~\ref{eq.ex6_sys}) with $\tau=0.15$, obtained using (a) the Galerkin approximation with \textit{ode15s} and (b) the \textit{dde23} solver (all tols.\ $10^{-6}$).}
    \label{fig.ex6_attractor2}
\end{figure}

\begin{figure}[t]
    \begin{center}
        \includegraphics[width=0.9\textwidth]{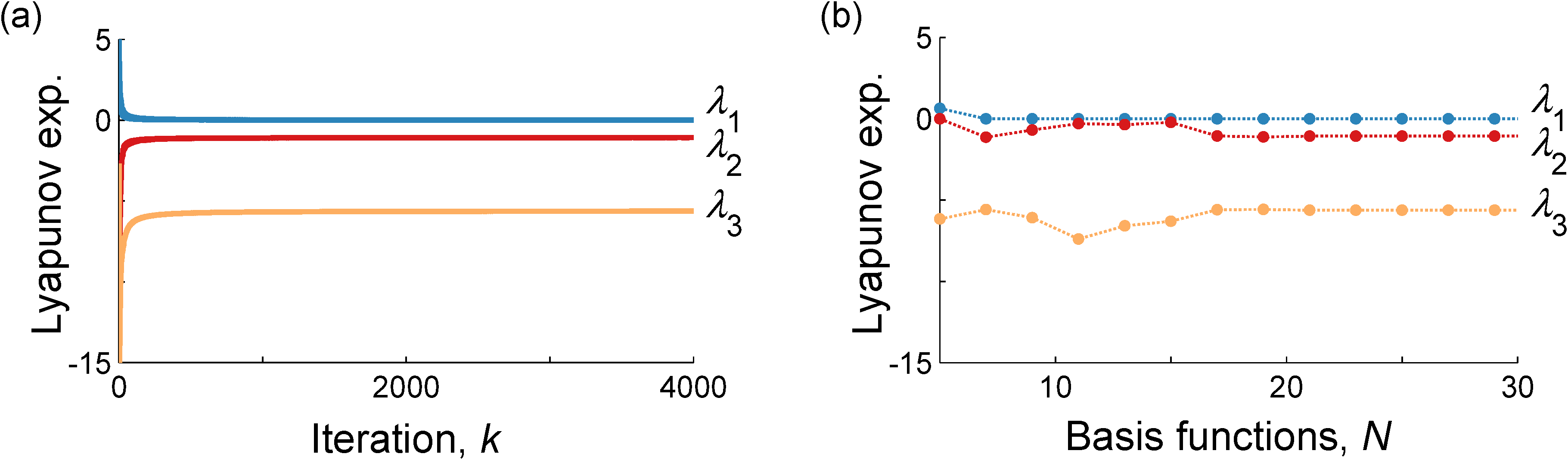}
    \end{center}
    \caption{Dominant Lyapunov exponents of the delayed Lorenz attractor (Eq.~\ref{eq.ex6_sys}) with $\tau=0.15$, computed using Galerkin approximation (\textit{ode15s}; tol.\ $10^{-6}$). The values of the Lyapunov exponents are shown (a) as $k$ increases for $N=20$ and (b) as $N$ increases.}
    \label{fig.ex6_lyapunov2}
\end{figure}

Lorenz derived a three-dimensional system from a 12-dimen\-sional model of atmospheric convection \cite{lorenz1963deterministic,saltzman1962convection}.
Lorenz found that the resulting system of ODEs was sensitive to initial conditions and exhibited chaotic behavior.
In this example, we modify the Lorenz system by introducing a delay as follows:
\begin{subequations}\begin{flalign}
    \label{eq.ex6_sys1}
    \dot{x}(t) &= \sigma \left( y(t) - x(t) \right)\\
    \label{eq.ex6_sys2}
    \dot{y}(t) &= \rho x(t) - y(t) - x(t) z(t-\tau)\\
    \label{eq.ex6_sys3}
    \dot{z}(t) &= x(t) y(t) - \beta z(t-\tau)
\end{flalign}\label{eq.ex6_sys}\end{subequations}
We use parameters $\sigma=10$, $\rho=28$ and $\beta=8/3$, and investigate two values of delay $\tau$.
In Fig.~\ref{fig.ex6_attractor1}, we show the attractor of Eq.~\eqref{eq.ex6_sys} from the Galerkin and direct solutions with delay $\tau=0.13$; the Lyapunov exponents are shown in Fig.~\ref{fig.ex6_lyapunov1}.
With $\tau=0.13$, the dominant Lyapunov exponent is positive and the system is chaotic.
The attractor and Lyapunov exponents with delay $\tau=0.15$ are shown in Figs.\ \ref{fig.ex6_attractor2} and \ref{fig.ex6_lyapunov2}, respectively.
With $\tau=0.15$, the dominant Lyapunov exponent is zero and the system is not chaotic.

\section{Conclusions}\label{sec.conclusions}
Lyapunov exponents are important indicators for detecting chaos.
A DDE is an infinite-dimensional system and, therefore, has an infinite number of Lyapunov exponents.
For practical reasons, DDEs are typically approximated by a finite-dimensional system of ODEs; the method of lines is a popular strategy for constructing this approximation.
In this work, we used a transformation to convert DDEs into PDEs with nonlinear boundary conditions.
These PDEs were converted into a finite-dimensional system of ODEs using Galerkin approximation; the tau method was employed for transforming the boundary condition.
Using Legendre polynomials as basis functions in the Galerkin approximation allowed us to generate closed-form expressions for the resulting ODEs.

Several examples were presented to demonstrate the efficacy of the proposed approach.
Galerkin approximation was shown to generate solutions with lower error than systems generated using the method of lines, even when the latter had substantially higher dimension.
If the proposed Galerkin approach is used, the standard algorithm for computing the Lyapunov exponents for systems of ODEs can be employed to compute the Lyapunov exponents for systems of DDEs.
We have demonstrated that the Lyapunov exponents converge as the number of terms in the Galerkin approximation ($N$) increases.
Finally, the attractors obtained using Galerkin approximation closely matched those obtained via direct simulation of the original DDEs.
In all examples, the Galerkin method produced reliable approximations of attractors and Lyapunov exponents using only 30--50 ODEs, suggesting that the Galerkin projection may lead to smaller---and, therefore, more computationally efficient---systems of ODEs than the conventional method-of-lines approach.

\section*{Acknowledgements}
C.P.V.\ gratefully acknowledges the Department of Science and Technology for funding this research through the Fast Track Scheme for Young Scientists (Ref:SB/FTP/ETA-0462/2012).

\bibliographystyle{asmems4}
\bibliography{SadathUchidaVyasarayani2018}

\end{document}